# Cold molecules: Progress in Quantum Engineering of Chemistry and Quantum Matter

John L Bohn,[1] Ana Maria Rey,[1] and Jun Ye[1]

[1]JILA, National Institute of Standards and Technology and University of Colorado Boulder, CO 80309-0440, USA
Email: bohn@murphy.colorado.edu, arey@jilau1.colorado.edu, ye@jila.colorado.edu

## Abstract

Cooling atoms to ultralow temperatures have produced a wealth of opportunities in fundamental physics, precision metrology, and quantum science. The more recent application of sophisticated cooling techniques to molecules, which have been more challenging to develop due to complex molecular structures, has now opened door to the longstanding goal of precisely controlling molecular internal and external degrees of freedom and the resulting interaction processes. This line of research can leverage fundamental insights into how molecules interact and evolve to enable the control of reaction chemistry and the design and realization of a range of advanced quantum materials.

Molecules hold a central place in the physical sciences. On the one hand, molecules consisting of a small number of atoms represent the upper limit of complexity we can at present hope to understand in complete detail, starting from quantum mechanics. On the other hand, molecules are the building blocks from which more complex phenomena emerge, including chemistry, condensed matter, and indeed life itself. These molecules then represent a kind of intellectual fulcrum around which we can leverage our complete understanding of small systems to probe and manipulate increasingly complex ones.

Precisely controlled studies of molecules started decades ago, with the invention of supersonic molecular beams for cooling (*1*) and coherent control for manipulation of internal states (*2*). However, bringing the temperature of molecular gases to the quantum regime is a relatively recent endeavor (*3*). When molecules move extremely slowly in the laboratory frame, and control of their internal degrees of freedom is achieved at the individual quantum state level, then each step of a complex chemical reaction can in principle be monitored and measured. The energy resolution underpinning such a process would be limited only by fundamental quantum rules that govern the molecular interaction from start to finish. The capability to track in full detail how multiple molecular species approach each other, interact via their evolving potential energy landscape, form intermediates, and re-emerge as final products, all while monitoring the internal and external energy level distributions, may have seemed out of reach only a few years ago. However, thanks to the recent progress in the field of cold molecules, we could soon be able to do precisely that. First-principle understanding of the most fundamental molecular reaction processes will furthermore enable the design and control of complex molecular transformations and materials with powerful functionality.

Molecules have rich energy level structures, owing to the vibrational and rotational degrees of freedom, compared to atoms. This presents a challenge for cooling technology. However, once we gain control over these molecular degrees of freedom, we create opportunities to take advantage of their unique properties, such as precise manipulation of long-range interactions mediated by molecular electric dipole moments. This capability will enable exploration of emergent collective phenomena in interacting many-body systems, which represents one of the central challenges in science. Besides studying chemical reactions in the new quantum regime, we may use cold molecules to synthesize quantum materials with strong correlations that could shed light on poorly understood phenomena such as superconductivity, quantum magnetism, and topological order.

Ultracold atoms have already played a revolutionary role in bridging the scientific quest between simple quantum systems and many-body physics. Synthetic materials assembled with ultracold atoms are typically billions of times less dense than electronic materials, with constituents many thousands of times heavier than electrons. These systems need to be cooled to nanoKelvin temperatures before collective quantum effects start to emerge, leading to rich phenomena qualitatively similar to those found in real electronic materials. Low energy scales can provide great advantages, such as observability in real time, but they also impose new challenges at the same time. We can overcome some of the challenges with polar molecules, as their strong and long-range interactions mitigate the requirement of cooling to ultralow motional energies.

Precise control of molecular states and the corresponding interactions is thus of paramount importance. Achieving this capability will allow us to understand chemical processes on the most fundamental, quantum mechanical bases, and thus facilitate control of both coherent and dissipative molecular interaction processes. Full molecule control will also help construct molecule-based synthetic quantum matter to study strongly correlated quantum phenomena. Our aim here is to provide this underlying connection between different intellectual pursuits based on cold molecules.

**Chemical physics.**

The theme of chemical physics is to know, understand, and eventually control how reactants become products. The idea of completely controlling and probing chemical reactions encompasses several goals, as illustrated in Fig. 1. First, reactant molecules should be prepared in individual internal quantum states, having for instance well-defined quantum numbers of electronic excitation, vibration, rotation, orientation, and alignment. In cases where it is relevant, control over electronic spin and perhaps even nuclear spin degrees of freedom of the reactants is also desirable. In addition, the collision itself should be initiated by manipulating the distribution of relative velocities of the colliding reactants. The second goal of control and probing is to measure the species constituting the products, as well as their populations in their own electronic, vibrational, rotational, and spin degrees of freedom. Relative velocities of the products are distributed according to differential cross sections, that is, angular distributions of products, which carry information about the interaction process. A final goal, and surely the most ambitious one, would be to observe, or even manipulate, the atomic complex formed during the reaction. A strong focus of this effort would be to identify transition states, in terms of their energies and atomic configurations; the barriers between them; and perhaps even the time evolution from one transition state to the next along the reaction path. Recent progress in achieving these goals has been substantial and in some cases astonishing, as we highlight below.

State preparation and readout are largely spectroscopic procedures and can therefore be accomplished with high precision. They are assisted moreover by the well-controlled velocity distributions produced by molecular beams. These abilities have led to a large and rich literature already on state-to-state chemical reaction studies, described in reviews and textbooks (*4, 5*). Under ordinary circumstances, these experiments resolve vibrational and rotational states, with tunable translational kinetic energy distributions at fractions of electron volts, suitable for probing reaction barriers.

It is not our intention to revisit this vast field, but rather to explore its extension to ever finer spectroscopic resolution and far lower translational energies. Cold molecules cover a range of temperatures usually from a few Kelvin to a few milliKelvin. Ultracold refers to the regime wherein collisional processes require explicit quantum mechanical treatment (*6*). The field of cold molecule studies is by now large and has already produced many clever insights and developments, of which we will only highlight a few by way of illustrating some main ideas. More comprehensive reviews are available for further studies, for example (*7-11*).

We therefore emphasize the situation where, beyond vibrations and rotations, fine structure and hyperfine structures are also resolved and can play a role in scattering dynamics. By considering translational temperatures on the sub-milliKelvin scale, research arrives in a novel regime where collision energies are the smallest energies of the problem. At the very lowest temperatures, effects are seen that resolve individual partial waves of scattering. Finally, the attainment of extremely slow molecules opens new opportunities for manipulating the initialization of chemical reactions, foreign to conventional molecular beams.

It may seem counterintuitive that chemistry can be a subject of study at all at temperatures that are orders of magnitude below typical reaction barriers. Nevertheless, the seminal (theoretical) work of Balakrishnan and Dalgarno (*12*) showed that the rate of a chemical reaction can be significant at zero temperature. This effect, studied first in the F + $H_2$ → H + HF reaction, is due to a van der Waals resonance near threshold, which facilitates tunneling through the reaction barrier (see below). At sufficiently low temperatures, when the deBroglie wavelength of relative motion of the reactants exceeds the range of their interaction, reaction rate constants become independent of energy, as described by the Wigner threshold law (*6*), rather than dropping as they do at higher temperatures, governed by the law of Arrhenius. Control of reactants is then a means of manipulating the way in which the reaction barrier is breached (*13*).

*Reactants*

Understandably, molecules are controlled most easily before the reaction begins. The reaction itself, involving nanometer length scales, electron-Volt energy scales, and sub-picosecond time scales, leaves little room for influencing the atoms *en route*. By contrast, manipulating molecular samples before they are sent off to react gives the experimenter the opportunity to do the necessary state preparation, which can have decisive consequences on reaction pathways and final products.

The techniques of molecular state preparation reside largely in exploiting and manipulating thermodynamics. For decades the workhorse of chemical physics has been the molecular beam, which exploits the cooling effect experienced by molecules that squirt through an aperture from a high-pressure enclosure into vacuum. This cooling process usually results in a population of molecules in very low-lying vibrational and rotational states. From here, the molecules can be state-selected or driven into the desired internal states by applied DC or radiation fields (*14*).

Molecular beams also have an impact on translational motion, producing a non-thermal distribution of velocities that has about a 10% width around its mean velocity at hundreds of m/s in the lab frame. To achieve better control over relative velocities of distinct reactants, it has long been common practice to employ two such beams in a crossed configuration. Colliding the beams head-on achieves large relative velocities, whereas beams running nearly parallel have fairly slow relative velocities, but a finite remaining angle limits the achievable energy resolution. The Narevicius group recently solved this long-standing problem by merging two beams perfectly in a curved magnetic guide (shown in Fig. 2A), achieving collision energy below 1 K (*15*).

Other techniques have emerged to slow and cool the beams in the laboratory frame. This may be done in the beam apparatus itself, for example, using the "effusive beam" technique developed by the Doyle group (*16*) wherein the beam emerges from a pressurized buffer gas of atoms that are already cold, with temperatures in the range 2 to 20 K. This pre-cooling mechanism, if coupled appropriately through an aperture, can produce a beam of molecules much slower, narrower, and better focused than a typical beam from a hot source.

Extracting a narrower subset of a beam's velocity distribution can further improve velocity resolution. Fortunately, many molecules that possess electric dipole moments respond to laboratory-strength electric fields. For example, a properly oriented ammonia molecule, moving from a field-free region "uphill" unto a region of space where the electric field is 100 kV/cm, could be brought to rest from a speed of 60 m/s. As realistic molecular velocities are higher, in practice multiple stages of electrodes, properly arranged in an array termed a "Stark decelerator," are used to bring a portion of the molecules to rest in the laboratory. This technique, pioneered by the Meijer group and extended by Lewandowski among others, has been used both as a controlled beam source (*17-19*) and as a means of loading molecules into electrostatic or magnetic traps, where the typical temperature will be hundreds of milliKelvin – a major reduction of relative velocities (*20, 21*). In a similar vein, electrodes of a curved geometry can serve to guide sufficiently slowly moving molecules away from the main beam, isolating them from the unwanted, faster molecules (*22*).

These techniques employ conservative electric potentials, and therefore conserve the total phase space density, shunting slow molecules to where they are useful and throwing away the rest. The samples tend therefore to be small, although collision studies can often be usefully done (*23*). For example, an influence on the collision cross section due to dipolar interactions has been observed when an electric field guided ammonia molecular beam (produced from a buffer gas cooled cell ) collided with a magnetically trapped sample of hydroxyl radicals (produced from a Stark decelerator) (*24*). The production of cold, comparatively high phase-space-density samples on the other hand does require an actual cooling mechanism. This requires manipulating thermodynamics to provide a dissipative mechanism.

One such mechanism exploits the fact that the electric forces that slow a molecule are different for different internal states of the molecule. Thus, if a molecule in a high-energy state climbs a potential hill and loses kinetic energy, that kinetic energy can be permanently removed upon driving the molecule to a low-energy state, thus introducing the dissipation. For correctly chosen field configurations, this process can be cycled, ultimately lowering the molecular temperature. This mechanism has been successfully demonstrated by the Rempe group, producing cold, trapped $H_2CO$ molecules (*25*). Another dissipative cooling process entails evaporation of trapped molecules. In the case of hydroxyl radicals (OH) trapped in a magnetic trap, a radio frequency "knife" was used to selectively remove hot molecules from a trapped to an untrapped state (*26*). The use of electric and magnetic fields together however leads to a severe Majorana spin flip loss mechanism (*27*), limiting the success of this approach until this problem is addressed.

A complementary technique, laser cooling, has been very successful for the field of cold atoms, it has been a long road extending to molecules. To bring a moving molecule to rest via laser cooling, the molecule needs to absorb and spontaneously emit tens of thousands of photons. However, upon the first spontaneous emission, the molecule may reside in any of a large number of ro-vibrational states, from which it can no longer absorb the second photon, thus ending the cycling process.

Fortunately, for a few selected classes of molecules it is possible to find an excited state that decays preferentially back to a useful ground state (*28*). Combined with proper angular momentum selection rules, a magneto-optical trap (MOT) for molecules was proposed by the Ye group (*29*). The DeMille group soon reported a demonstration by laser-cooling SrF molecules (*30*). This was followed by construction of a two-dimensional magneto-optic compression of YO molecules in the Ye group (*31*) and a 3D MOT of SrF in the DeMille group (*32*). Other molecular species have recently been laser cooled as well (*33, 34*), particularly exciting is the achievement of large numbers of CaF molecules captured in 3D MOT (*35, 36*).

Molecules can now be confined in traps constructed from spatially inhomogeneous magnetic, electric, or optical fields including far-detuned optical dipole traps or optical lattices. Typically, the samples need to have temperatures on the order of 1 to 100 milliKelvin to be confined by laboratory magnetic or electric fields, and ultralow temperatures of a few tens of µK and below for optical traps. Trapping molecules brings a completely new perspective to collisions. First is the substantial reduction in relative velocities, down to 0.01 to 1 m/s in favorable cases, which promises sufficient energy resolution to observe scattering resonance in chemical reactions. Second, there is a shift in perspective relative to the traditional beam approach, as the trapped molecules travel in all directions, eliminating in most cases a fixed collision axis. The concept of an impact parameter, the closest approach $b$ between two classical trajectories in the absence of their interaction, is replaced with that of a partial wave, which better represents the quantum-mechanical realities of the situation.

Moreover, in the low energy regime achieved in traps, quantum mechanics restricts angular momentum to integer multiples of the reduced Planck constant $\hbar$. The relative angular momentum $L=\mu v b$ is conserved, where $\mu$ is the reduced mass of the colliding pair and $v$ is their relative velocity. Thus, very roughly speaking, the available impact parameters are $b = \hbar l/\mu v$, where $l = 0, 1, \ldots$ is used to index the partial wave of the collision. For zero angular momentum, "head-on" collisions with $b=0$ are thus possible, and chemistry can occur. However, if the molecules possess any nonzero relative angular momentum, they are kept apart by the centrifugal potential, and the distance of closest approach increases as the energy is reduced. The question of how close together the reactants must be to initiate a possible reaction depends on the value of the constant $C_6$ in the long-range intermolecular interaction $-C_6/r^6$, where $r$ is the distance between reactant molecules; this in turn defines an energy below which a given partial wave $l$ will not contribute to scattering. For a light, weakly polarizable species, such as NH, this energy scale is on the order of 10 milliKelvin (*37*). Below this temperature, the colliding molecules are generally guaranteed to have no angular momentum about each other.

An ultimate goal in molecular collisions would be to extract the scattering from each partial wave – a discrete quantity – rather than having to settle for observables averaged over a continuum of impact parameters. This goal was realized in ultracold KRb molecules produced in the Jin /Ye groups (*3, 38-40*). In this case molecules were welded together optically from ultracold K and Rb atoms, ensuring a translational temperature of the molecules of ~100 nanoKelvin, with the molecules in a single internal quantum state, including the absolute lowest energy state accounting for nuclear spins. These molecules are presumably susceptible to the barrierless reaction KRb + KRb → $K_2$ + $Rb_2$, which manifested as loss in the gas, although the products were not observed.

As identical fermions in identical nuclear spin states, these molecules were forbidden by quantum mechanical symmetries from scattering via $l = 0$ partial waves, and so scattered according to angular momentum $l = 1$. However, on changing the nuclear spin for half the molecules, this symmetry no longer held, and the molecules could again collide head-on with $l = 0$. Because there was no centrifugal barrier associated with this channel, the reaction rate was enhanced by almost two orders of magnitude. This was therefore the first instance of choosing the partial wave degree of freedom in relative motion, thus completing the task of controlling the reactants at all possible degrees of freedom of the collision. Further control of the molecules could be implemented by engineering of the spatial configuration of the trap to restrict molecular motions in low spatial dimensions (*41*) and by introducing external fields to control the molecular orientation and dipolar moment in the lab frame (*42*). These experiments pave the way for manipulating polar molecules in an optical lattice to form synthetic quantum materials, as discussed in the second half of this Review.

*Products*

Having prepared the reactants, it is necessary to observe and perhaps even control the products of reaction. This development, too, has a long history in beam experiments. The principal ambition of product investigation is to ascertain which chemical species emerge, in which internal states, and in which direction they are moving. The first two factors are often accessible via standard spectroscopic techniques, such as laser-induced fluorescence.

To measure the velocities of outgoing products, techniques such as velocity map imaging (VMI) are employed (*43*). An outgoing molecule is first selectively ionized with little perturbation of its momentum. An electric field then guides the ion to a multichannel plate, which records the location of its arrival. The mass and charge of the ion and the shape of the electric field connect the original velocity of the molecule to the impact location on the plate.

To capitalize on the velocity distribution and extract the angular distribution of products requires that the velocities be referred to a well-defined incident axis of the collision. VMI techniques are thus most useful when the incident beams are as well-controlled as possible, using techniques described above. This type of control was exhibited, at least for inelastic scattering, by the van de Meerakker group, who crossed a pulsed-beam with a Stark decelerator. This experiment succeeded in measuring narrow diffraction oscillations in the differential cross sections of NO scattering from rare-gas atoms (*44*).

In the coldest of samples, residing in traps, measurements of differential cross sections are problematic precisely because there is no well-defined initial direction for collisions, as the molecules are traveling in many different directions. Still, identifying the products and their distribution into internal states can prove valuable. Many trapped-molecule experiments are not well suited to measuring products, at least currently. Part of the problem is that the products of reaction can emerge with large kinetic energies, on the order of electron-Volts (corresponding to many thousands of Kelvin), and are therefore not themselves trapped for further study. At least two types of experiments can circumvent this issue, however.

The first of these was recently demonstrated by the Weinstein group, studying the reaction Li + CaH -> LiH + Ca (*45*). This experiment was not conducted in a trap, but in a helium buffer-gas cell at a few Kelvins, where both the reactants and products were stored. Despite releasing 0.9 eV of energy in the reaction, the products quickly cooled to the ambient temperature and were available for interrogation. Rotational levels cooled quickly and so their nascent distribution after reaction was undetermined. However, vibrational relaxation is notoriously slow in a buffer gas, meaning that vibrational populations produced from the reaction were still available for viewing.

A second option for harnessing the products is provided by ion traps. Due to the comparatively strong electromagnetic field confinement, molecular ions can be trapped even after they react, if the product is also electrically charged. Remarkably, ions in such a trap arrange themselves into a "Coulomb crystal," where they can be individually observed by fluorescence. When a slowed reactant beam flows through the crystal, reactions can be observed one by one as the ions vanish from the crystal, providing the ultimate number resolution of a chemical reaction.

If the products happen also to be ionic, they can be sympathetically cooled by the original ions, and join the Coulomb crystal themselves. This occurs, for example, in the experiments of the Softley group (*46*), where certain calcium ions would disappear from the crystal during the reaction $Ca^+ + CH_3F \rightarrow CaF^+ + CH_3$. The $CaF^+$ ions were also trapped, and made their presence known by perturbing the $Ca^+$ crystal, even though $CaF^+$ was not directly observed. Knowing it is there, though, it is ripe for exploration.

*Transition States*

Careful preparation of reactants and observation of products often yield no progress toward the even greater challenge of observing or manipulating intermediates during the reaction. Generally, the task of following the reaction in transition is the job of theory. However, the reaction itself contains useful handles by which it can be manipulated. These handles appear in the form of resonant states or transition states of the collision complex. That is to say, if reactants A and B (each of which may consist of many atoms) collide, they may form various configurations of all the atoms in the AB cluster, before finding their way to the products of reaction. This situation can lead to a resonant enhancement of the cross section at the energy of the AB cluster.

One such set of cluster states comprises the van der Waals states, held together by the weakly attractive van der Waals interaction between molecules, but prevented initially from reacting by an energetic barrier (*47*). More intricate resonant states, termed transition state resonances, may exist behind the barrier. Observing and characterizing these resonances can shed light on the way in which the atoms share energy and pass it between them, revealing hints about the chemical process. Cold, controlled collisions of reactants may allow the energy resolution necessary to observe these resonances. Moreover, if the resonant states happen to have magnetic or electric moments different from those of the reactants, then it is possible to apply a corresponding field to scan one or more resonances through the collision energy. This is a process long known and extremely useful in ultracold scattering of atoms (*48*). For chemistry, its power is only starting to be appreciated, for example in recent detailed calculations of H + LiF -> Li + HF reactions by Tscherbul and Krems, who predict tremendous variation of rotational product states when driving electric field resonances in this system (*49*).

Another recent, striking example shows the interplay of theory and experiment in such cases. This study involved collisions of $H_2$ molecules with excited (metastable) helium atoms, leading to an ionization process whose products were easily detected (*50*). The potential energy surface that dictates this reaction is dependent on the angle θ between the axis of $H_2$ and the line joining the center of mass of $H_2$ to the He atom.

This angular dependence was probed resonantly in the experiment, as follows. The potential energy is generally written as an expansion into Legendre polynomials, $V(R,\theta) = V_0(R)P_0(\cos\theta) + V_2(R)P_2(\cos\theta)$. If the molecule was prepared in a rotationless quantum state, with rotation quantum number $j=0$, then the collision could not experience the anisotropy afforded by $P_2$, and a featureless cross section resulted from scattering due to the isotropic component $V_0$. However, preparing the molecule in a $j=1$ state effectively allowed the collision to probe the complete potential, including anisotropy and a resonant state (Fig. 2B). The determination of the resonant energy places a tight constraint on the calculations of the potential energy surface. As pointed out by Simbotin and Côté, due to molecular symmetry, preparing the $H_2$ molecule in $j=0$ or $j=1$ is a matter of selecting its spin state (*51*). Thus, this reaction is another case in which reactions taking place at eV energies are controlled by nuclear spin degrees of freedom that is associated with a much smaller energy scale.

Finding resonances in scattering can thus be informative, but serious theory is needed to render interpretation useful. This is not a drawback, as the ultimate goal is in fact interpretation. Still, as an alternative, direct spectroscopic probes of transition states are desirable. The power of spectroscopy was demonstrated, for example, in the benchmark chemical reaction F + $H_2$ -> HF + H (*52*). Here, the experiment began with ground-state clusters of all three atoms, with an electron attached. From these $FH_2^-$ clusters, the electron was easily detached by light absorption, yielding resonances that could be attributed to transition states of the neutral $FH_2$ complex. As a result, the energetics and identity of the transition states were amenable to study.

After capturing the transition states, the ensuing goal is to understand the dynamics that drives the atoms through them, from reactants to products. Such a task would seem to require the spectroscopic accuracy of measuring the transition states, coupled with the time resolution to determine their populations in real time. In recent years such a tool, with broad spectroscopic capabilities and exquisite time resolution, has indeed been developed in the form of infrared frequency comb spectroscopy (*53*).

The power of this tool was unleashed very recently, in the atmospherically significant reaction OH + CO -> H + $CO_2$ (*54*). This reaction is known to proceed by forming a transition state HOCO, in both the *trans* and *cis* forms, which must surmount an energetic barrier before it can proceed to the products. While flopping around in this transition state, the complex can be affected in one of two ways by collisions with molecules in its environment. These collisions can either relax the complex into a lower energy state, stifling the reaction; or else activate it to higher energies, promoting its passage across the barrier. Watching the population of the state as a function of time, revealed the relative rates of successful to unsuccessful reactions as a function of the background pressure.

An important challenge associated with the pursuit of full quantum control over molecular processes is the breadth of a technique's applicability with respect to the full menagerie of interesting molecular species. Whereas some techniques, such as laser cooling, do require very specific optical sources for each molecule, many techniques successfully maintain much greater generality, including buffer gas cooling, molecular decelerators, and frequency comb spectroscopy.

*Theory and Prospects*

Amid the slew of experimental developments bringing new and detailed information on chemical reaction dynamics, the role of theory remains central to the ultimate goals of interpretation. The basic theoretical tools--construction of potential energy surfaces and scattering calculations based on those surfaces--are well-established, as detailed in a vast literature. The reader is again referred to Ref. (*7*) for an excellent survey in the context of cold molecules.

These calculations evolve in tandem with experimental data of ever-increasing quality, aided by increasing computational technology, but also by insights emerging from theoretical study. As an example of the latter, Cui and Krems have recently explored information-theoretic approaches to extract complex potential energy surfaces from a limited set of calculations (*55*). Clever insights like these can substantially reduce the computational effort required to provide understanding.

In the longer term, a vast array of experimental and theoretical avenues is closing in on the ultimate expression of what chemical reactions can reveal. Consider that in the end, quantum mechanics imposes discreteness on everything: molecules have well defined states in all internal degrees of freedom, from electronic structure all the way down to nuclear spin. Moreover, in the ultracold limit, even the relative motion of the reactant molecules becomes quantized into discrete partial waves, as was exploited in the KRb experiment referred to above.

Therefore, considering notation $\alpha$ to stand for the collection of quantum numbers of all reactant states, and $\beta$ that for all product states, the prospect of constructing the entire scattering matrix $S_{\alpha\beta}$ can be envisioned. This would constitute the "complete chemical experiment," whose analog in inelastic electron-atom or atom-atom scattering has been contemplated for decades (*56*). Chemistry affords the opportunity to go even beyond this, when reaction intermediaries and transition states are also observed. This activity would require the most stringent cooperation of experiment and theory, and would presumably teach us the most that Nature permits us to know about a given reaction.

## Quantum Materials

The realization of atomic Bose-Einstein condensation (BEC) set the stage for the creation of a new form of materials – synthesized quantum matter – that have stimulated insights into the development of many-body quantum correlations and fostered intimate connections between AMO and condensed matter physics (*57*). The remarkable capability of experimental control in nearly all possible aspects of atomic quantum gases facilitated the studies of BEC properties out of equilibrium, such as the dynamics of vortices and solitons (*58*). The production of atomic degenerate Fermi gases (*59*) opened a new window to the exploration of outstanding scientific questions such as superconductivity and superfluidity (*60, 61*), as well as tabletop studies of strongly correlated quantum fluids analogous to those formed in neutron stars and nuclear matter under strong interactions. In parallel, loading atoms in an artificial periodic potential generated by laser light, or optical lattices, have allowed the realization of simplified crystals with highly tunable geometry, degree of disorder, and strength of interactions. This development has led to many spectacular observations (*62*), such as the transition from a weakly-interacting fermionic or bosonic gas to a Mott insulator, a strongly correlated state where the large interatomic interactions prevent atomic motion and favor the formation of a structure with fixed number of atoms per lattice site.

Building on these great successes, to further advance our understanding of complex quantum materials and to provide insights to dynamical processes arising from strong correlations, important obstacles need to be overcome in this field. In fact, many key phenomena of crucial importance in condensed matter physics remain difficult to achieve and probe in state-of-the-art cold atoms experiments. Take the example of quantum magnetism, the fundamental origin of which features many open questions. Magnetic behavior in electronic materials emerges as a consequence of intricate interplays between quantum statistics, spin-orbit coupling and motional effects, as well as electromagnetic interactions. Magnetic correlations are often believed to be closely connected with unconventional superconductivity (*63-65*), and are at the heart of a large class of topological states of matter that are beyond the conventional description of the "Landau paradigm" of symmetry breaking (*66*). Generating non-local spin-spin interactions in a generic cold atom system is challenging since their contact interaction is short-ranged and magnetic dipole interactions are weak. When mediated with motion, magnetic correlations emerge only when the motional energy is less than the effective spin-spin coupling energy. Polar molecules, on the other hand, possess strong and long-range interactions with an enlarged set of internal states that are available to offer a more versatile platform for building synthetic quantum matter (*9, 10, 38-40, 67-70*). The basic question then is whether one can develop a quantum system of molecules that features precise quantum control at the same level as that demonstrated in atomic quantum gas experiments. While the front of cold molecules has been fairly broad using a diverse set of experimental approaches as discussed in the first half of this Review, the path towards the production of a quantum gas of polar molecules has been significantly more focused. In the remainder of this Review we will describe the only successful approach so far based on bialkali polar molecules.

We note that there are other AMO systems currently under experimental investigation, including trapped ions (*71*), Rydberg atoms (*72, 73*), and magnetic atoms (*70*), and they all offer the prospect of long-range interactions and a set of alternative attributes. In the current survey, however, we focus on polar molecules.

## *Quantum degeneracy of a molecular gas*

Polar diatomic molecules are excellent candidates for the investigation of magnetic phenomena. On one hand, they possess a permanent dipole moment that can be manipulated with external fields, and which provides long-range anisotropic dipolar interactions. On the other hand, they exhibit a hierarchy of internal degrees of freedom including hyperfine, rotational, vibrational and electronic levels. Once controlled, the rich internal structure provides a great degree of flexibility and specificity to engineer a quantum system by using external fields to couple molecules directly via dipolar interactions without the need of motional processes. A fundamental requirement to take full advantage of these attributes is the capability to prepare a molecular quantum gas.

As discussed in the first half of this Review, cooling molecular gases to the quantum regime is extremely challenging. In fact, none of the techniques discussed so far for direct cooling of molecules have come close to the quantum degenerate regime: most have produced molecular phase-space densities that are about 10 – 12 order of magnitude lower than those required for quantum degeneracy.

In 2008, a revolutionary path emerged for successful production of an ultracold polar molecule gas near quantum degeneracy (Fig. 3). In this approach, ultracold polar molecules were coherently assembled from two atomic species that had themselves been brought to quantum degeneracy. Specifically, the Jin/Ye groups employed a Fano-Feshbach resonance (*48*) for magneto-association of pairs of fermionic $^{40}$K and bosonic $^{87}$Rb atoms (*3*). These weakly bound (highly vibrationally excited) Feshbach molecules were then coherently transferred to the absolute rovibrational ground state in the ground electronic potential using a pair of phase-coherent lasers coupled to a common intermediate electronic excited state. This population transfer process, known as stimulated Raman adiabatic passage (STIRAP), is fully coherent in that nearly 100% of the Feshbach molecules were transferred to the single ground state while imparting hardly any heating to the molecular gas. (STIRAP was used also to create ground-state homonuclear $Cs_2$ molecules (*74*).) Once in the rotational-vibrational-electronic ground state, molecules were transferred to the hyperfine ground level using microwave fields that couple the ground and the first excited rotation states (*75, 76*). The net result was a gas of fermionic $^{40}$K$^{87}$Rb molecules at its lowest internal energy state with a temperature of ~100 nanokelvin and a phase-space density just around the onset of quantum degeneracy. Expressed in terms of the Fermi temperature, $T_F$, the molecular gas was about $T/T_F$ ~1.5. Further improvements in this method have recently led to the production of a quantum gas of $^{40}$K$^{87}$Rb loaded in a three-dimensional optical lattice, with entropy per particle corresponding to that of a bulk Fermi gas with $T/T_F$ ~ 0.3 (*77*).

In recent years (see Fig. 3), the production of rovibrational ground state molecules has also been achieved for several other bi-alkali species, such as $Cs_2$ (*74*), RbCs (*78, 79*),

NaK (*80*), and NaRb (*81*). This effort is motivated partly by the desire to avoid the exothermic reaction as experienced by KRb molecules (*82*), and partly by the appeal of larger dipole moments (*83*). However, it is possible to have molecular loss via a three-body collision process, during which two molecules collide and temporarily form a reaction complex when a third molecule approaches sufficiently closely and force inelastic loss of all three molecules (*84*). With a number of groups worldwide actively working on polar molecule quantum gas experiments, we can expect that some of these open questions will be addressed and a diverse set of quantum gases of polar molecules will become available in the near future.

*Dealing with reactive losses*

While the original motivation of the KRb experiment for the creation of a molecular quantum gas was to study correlated quantum matter, the experiment turned into a unique opportunity to study chemical reactions near absolute zero, where simple quantum mechanical rules govern how single-state-controlled reactants approach each other. The two-body loss evident from a gas of ground-state KRb molecules confined in an optical dipole trap was attributed to an exothermic reaction process of KRb + KRb $\rightarrow$ $K_2$ + $Rb_2$, leaving the products with sufficiently high kinetic energies to almost instantaneously escape the optical trap (*6, 13*). For the lowest collision partial wave ($l = 1$), the associated centrifugal barrier of 24 µK was significantly higher than the collision energy of ~ 100 nK. Thus, the reaction was largely suppressed, leading to a lifetime of ~1 s for a molecular density of $10^{12}$ cm$^{-3}$ in the trap.

When an electric field is applied, molecules become oriented and exhibit a dipole moment in the lab frame. Dipolar interactions mix different partial wave collision channels, substantially altering the two-body collision process (*13, 42*). In fact, the good quantum number to use is the projection of the angular momentum along the external field: when two molecules collide in a plane perpendicular to their oriented dipoles, the collisional energy barrier is increased; when they collide along their orientation axis then the barrier is lowered. For this reason, when both 'side-to-side' and 'head-to-tail' collisions were allowed under an applied electric field, the observed lifetime was reduced to a few milliseconds at typical molecular densities of $10^{12}$ cm$^{-3}$ in the trap (Fig. 3).

The control of dipole-mediated chemical reactions was demonstrated by suppressing the undesirable 'head-to-tail' collisions via tight confinement in geometries that only energetically allow repulsive interactions. For example, by trapping molecules in an array of quasi-2D disks generated by interfering an optical beam (in the so-called 1D optical lattice) and aligning the dipoles by an external electric field perpendicular to the disks, chemical reactions were observed to slow down with increasing dipole moment (*42, 85, 86*). This experiment illustrates the use of optical lattices and electric field manipulation to render the molecules more stable for condensed matter purposes.

A tightly confining three-dimensional optical lattice freezes out molecular motion in all directions (Fig. 3). Trapped in individual lattice sites each molecule shows a lifetime longer than 20 s, limited only by off-resonant light scattering from the trapping beams (*87*). The strong inter-molecular interactions, both reactive and conservative, guarantee that only one molecule can occupy a particular lattice site. Here, another intriguing quantum phenomenon manifests in the overall reaction rate of the molecular gas, the so-called quantum Zeno effect. The molecular loss rate is increasingly suppressed with an enhanced onsite reaction rate as a result of the tightened spatial confinement (*88, 89*). When the onsite reaction rate far exceeds the tunneling rate, hopping of molecules between neighboring sites becomes a second order process that is increasingly suppressed.

*Synthetic quantum magnets*

The preparation of a stable, low entropy gas of polar molecules pinned in a 3D lattice set the stage for exploration of quantum magnetism mediated by dipolar interactions (Fig. 3). The phrase magnetism refers to the modeling of an array of coupled spin-½ magnetic moments with an interacting spin-½ system where couplings are mediated by electric dipolar interactions (*38-40*): If a molecule located at site *i* with a dipole moment $\hat{\boldsymbol{d}}_i$ interacts via a dipolar interaction with another molecule at site *j* with a dipole moment $\hat{\boldsymbol{d}}_j$, the interaction is given by

$$\hat{H}_{ij}^{dd} = \frac{1}{4\pi\epsilon_0}\left(\frac{\hat{\boldsymbol{d}}_i\cdot\hat{\boldsymbol{d}}_j - 3(\hat{\boldsymbol{r}}_{ij}\cdot\hat{\boldsymbol{d}}_i)(\hat{\boldsymbol{r}}_{ij}\cdot\hat{\boldsymbol{d}}_j)}{|\boldsymbol{r}_i-\boldsymbol{r}_j|^3}\right), \quad (1)$$

where $\boldsymbol{r}_{i,j}$ are the vectors describing molecular positions *i* and *j*, and $\hat{\boldsymbol{r}}_{ij} = \frac{\boldsymbol{r}_i-\boldsymbol{r}_j}{|\boldsymbol{r}_i-\boldsymbol{r}_j|}$ is the unit vector joining the molecules. If the molecules are prepared and controlled to populate only two opposite-parity rotational states, denoted as $|\uparrow\rangle$ and $|\downarrow\rangle$, then one can express the dipole-dipole interaction Hamiltonian in terms of spin-½ angular momentum operators ($\hat{S}$) acting on the $\{\uparrow,\downarrow\}$ states. The resulting Hamiltonian, $\hat{H}^S$, represents an XXZ spin-½ system, which is an iconic model for quantum magnetism,

$$\hat{H}^S = \sum_{ij}\hat{H}_{ij}^S,$$

$$\hat{H}_{ij}^S = V_{dd}(\boldsymbol{r}_i - \boldsymbol{r}_j)\left[J^\perp\left(\hat{S}_i^+\hat{S}_j^- + \hat{S}_i^-\hat{S}_j^+\right) + J^z\hat{S}_i^z\hat{S}_j^z + W(\hat{S}_i^z + \hat{S}_j^z)\right], \quad (2)$$

with $V_{dd}(\mathbf{r}_i - \mathbf{r}_j) = (1 - 3\cos^2(\Theta_{ij}))/|\mathbf{r}_i - \mathbf{r}_j|^3$ and $\Theta_{ij}$ the angle between the quantization axis, set by the external electromagnetic field, and $\hat{\mathbf{r}}_{ij}$. The long range and anisotropic coupling constants, $J^\perp, J^z, W$, are fully determined by $\langle\sigma|\hat{\mathbf{d}}|\sigma'\rangle$ and can be tuned via the strength of the external field, or via the choice the rotational state. Whereas $J^z$ (Ising interaction) and $W$ vanish under zero electric field, $J^\perp$ (exchange interaction) remains finite and is intrinsically related to the transition dipole moment between the chosen pair of rotational states.

The first experimental implementation of $\hat{H}^S$ was realized in 2013 (*89*) using a 3D optical lattice sparsely filled with $10^4$ KRb molecules (~5%). The molecules were initially prepared in the rovibrational ground state $|\downarrow\rangle = |N=0, m_N=0\rangle$, where $N$ is the principal rotational quantum number and $m_N$ is its projection onto the quantization axis set by an external magnetic field. The hyperfine interaction lifts the degeneracy of the $N=1$ rotational manifold, consequently either $|N=1, m_N=-1\rangle$ or $|N=1, m_N=0\rangle$ could be selected as $|\uparrow\rangle$ by a microwave field to couple with $|\downarrow\rangle$. Under a zero DC electric field, only the $J^\perp$ term contributes to the dynamics by flipping spins between pairs of molecules in opposite spin configuration, $|\uparrow\downarrow\rangle \leftrightarrow |\downarrow\uparrow\rangle$. After preparing a 50-50 coherent superposition, the system was allowed to freely evolve under $\hat{H}^S$ for some time, after which the collective spin coherence was probed with the application of a second microwave $\pi/2$ pulse, followed with a spin population readout.

Of course, precise control of single molecule quantum states played an important role for the study of many-body physics. The real part of the molecular polarizability gives rise to the ac Stark shift that helps trap molecules in the lattice. The polarizability can be precisely controlled by varying the angle between the polarization of the lattice light and the quantization field defined by an applied electric or magnetic field. At a specific angle the polarizability of the two rotational states used as $|\uparrow\rangle$ and $|\downarrow\rangle$ becomes equal, thus minimizing the differential ac Stark shift between the two spin states, leading to prolonged spin coherence times (*90*). We note that a coherence time of ~ 1s was recently achieved between two hyperfine states of NaK in its ground rotational state (*91*).

The molecular spin-½ Hamiltonian resulted in clear experimental signatures for interacting many-body dynamics mediated by dipolar exchange interaction. The spin evolution fringe shown in Fig. 4A includes small amplitude oscillations at the frequencies determined by the exchange rate of rotational excitations between pairs of nearest-neighbor and next-to-nearest-neighbor molecules. The existence of several oscillation frequencies, arising from interactions of molecules separated by a few discrete distances in the lattice, results in an overall exponential decay of the fringe signal with a rate that is proportional to the lattice filling fraction. A specific dynamical decoupling pulse sequence can be designed to disentangle pairs of molecules under spin exchange, and thus suppressing the oscillation. All of these experimental observations have been supported by numerical calculations based on the $\hat{H}^S$ Hamiltonian (*92*).

Similar types of exchange processes have been observed using bosonic $^{52}$Cr atoms prepared in a Mott insulator state in a 3D lattice (*93, 94*). In those experiments, the spin was encoded in hyperfine states of the atoms and coupled by magnetic dipolar interactions. In the deep lattice limit when motion is frozen, the dynamics are described with a similar XXZ model. In contrast to the spin-½ KRb experiment, each Cr atom encodes a spin-3. For this system, the average populations for different spin sublevels and the total gas magnetization are not locked, and therefore the spin exchange dynamics can manifest directly in the spin population without observing spin coherence as was the case for KRb (Fig. 3b). Although magnetic interactions are generically weaker than electric dipolar interactions, the use of a smaller lattice constant under a higher filling fraction enabled the observation of the magnetic induced spin exchange dynamics in the $^{52}$Cr experiment. Moreover, the experiment observed the interplay between tunneling, onsite contact interactions and dipolar exchange magnetism (*93*).

This interplay has also been observed in a recent experiment with spin polarized $^{168}$Er atoms in a 3D lattice (*95*) (Fig. 3c). The long-range and anisotropic magnetic dipolar interactions introduced significant modifications to the superfluid-to-Mott insulator transition. For example, the critical point of the phase transition was observed to depend on the magnetic field direction and the aspect ratio of the trapped Er atomic gas.

Recently, another important step towards the creation of a low entropy gas of polar molecules in a lattice was accomplished via the development of a quantum synthesis protocol (Fig. 3) (*77, 96*). The idea was to create dual atomic insulators in the same optical lattice, a Rb Mott insulator, and a K band insulator, to maximize the number of individual lattice sites populated with exactly one atom from each species. This experiment required solving a number of challenges arising from the opposing requirements of the Bose and Fermi statistics associated with the two species, as well as their distinct masses and polarizabilities and the need to control the interspecies interactions during the state preparation and the magneto-association process. The protocol yielded a filling fraction of >25% of ground-state KRb molecules in the 3D lattice (*77*), corresponding to an entropy per molecule of just $2k_B$. The filling has reached the percolation threshold, where no isolated patch of molecules may exist in the lattice without feeling interactions sufficiently strong for correlations with the rest of the system. A variant of this protocol was recently used to produce low-entropy samples of RbCs Feshbach molecules (*97*). Future experimental improvements based on this work should be able to increase the filling further to 50% (*98*).

*Future prospects: Advanced materials*

The successful observation of dipole-induced quantum magnetism even in a sparsely filled molecular lattice is a manifesto of exciting physics waiting to be explored. Reaching higher lattice fillings, improving imaging resolution, and implementing stronger electromagnetic field control of the coupling constants will most likely make this strongly correlated quantum system computationally intractable with classical technologies. This quantum simulator could allow exploration of complex, non-equilibrium spin dynamics, and tracking of the propagation of quantum correlation, thermalization, and build-up of entanglement in long-range interacting spin systems (Fig. 5, bottom) (*99*). This would represent the first experimental system where the spatial dimensionality and the scaling of the range of interactions with respect to the interparticle distance are the same, making it a unique benchmarking system where theory is so far intractable. Moreover, empty sites in the lattice could be used as defects; hence we could use this platform for studies of spin transport and many-body localization (MBL) by tuning the molecular filling and dipolar interaction strength (*100*). A system displaying MBL transports neither heat nor charge (where by "charge" we mean "mass"), even when the amount of energy injected is macroscopic. MBL is currently of great interest, in part because a system displaying MBL may be used as a robust quantum memory or for implementation of topological order.

The thermodynamic phase diagram of $\hat{H}^S$ featuring effective long-range and anisotropic interactions is intractable with current classical computers. This type of model falls under the class of strongly frustrated systems, where competing interactions can prevent classical ordering even at zero temperature, exhibiting a magnetic analog of liquid phases (*101*). Such quantum spin liquids are highly entangled and can exhibit non-trivial topological behavior similar to the one found in the fractional quantum Hall effect including fractionalized excitations and robust chiral edge modes. The elucidation of phase transitions and diagrams using ultracold molecules would be a major advance in our understanding of strongly interacting systems (*102*).

Involving a larger set of rotational states in molecules would enable exploration of the emergence of spin-orbit coupling (SOC) in pinned dipoles (Fig. 5 middle). Spin–orbit coupling in solids is a key to understanding a variety of spin-transport and topological phenomena, such as Majorana fermions and recently discovered topological insulators. Implementing and controlling spin–orbit coupling in a synthetic quantum material is thus highly desirable. Despite great advances in using alkali-metal atoms to realize SOC (*103-105*), those implementations have been hindered by heating from spontaneous emission, which have limited the observation of many-body effects. For polar molecules, SOC can naturally emerge without the need for external laser fields since it is a result of dipolar-interaction-induced transfer of angular momentum between the internal rotational states. It manifests as spin excitations that carry a non-zero Berry phase. Recent theoretical studies have found that depending on the number of degenerated rotational states and the geometry of the molecular lattice the excitations can exhibit band structure similar to chiral excitations in bilayer graphene (*106*) or to Weyl quasiparticles (*107*). Experimental observables would include the density profile, spin currents, and spin coherences.

Dressing of molecules with microwave or optical fields leads to a very rich landscape of possibilities when multiple rotational states are involved. This enables the design of models in which the strength of the spin couplings depends on the spatial direction, predicted to lead to symmetry-protected topological phases (*108, 109*) or phases with true topological order such as fractional Chern insulators (*110, 111*).

The next phase of research may target several additional features of magnetism in real materials (Fig. 5 top). For example, magnetic interactions in real materials result from a complex variety of mechanisms, including exchange, super-exchange, and itinerant particles, occurring both in combination and in competition. An iconic example is the so-called *t-J* model (*112*), which is believed to contain the essential ingredients (tunneling, density–density and spin–spin interactions) to describe the high temperature superconducting state that emerges when a Mott insulator is doped. Although deceivingly trivial in mathematical expression, the model contains complex physics and is currently intractable with classical methodologies. If the temperature of a molecular quantum gas is lowered further, and the chemical reaction is controlled at the desired level, then the exchange interactions in the long-range analog of the *t-J* model will manifest rich dynamics in a dipolar gas (*113*).

Reaching a higher phase space density required for the observation of itinerant magnetism might be possible by evaporatively cooling the current molecular quantum gas. This requires tight confinement of the molecular gas into quasi-two-dimensional traps, stable under a strong DC electric field (*85*). Most of these capabilities are under experimental development. An intriguing direction that evaporative cooling will open in molecules prepared in the same internal state in a stack of two-dimensional traps is the formation of dipolar chains and clusters extending over multiple trap sites (*67, 68, 70*). Those structures are stabilized by the intra-layer attraction. In the presence of appropriate microwave dressing, intra-layer dipolar attraction can also stabilize a topological superfluid $p_x + ip_y$, a target for topologically protected quantum information processing (*68*).

**Figures and Captions:**

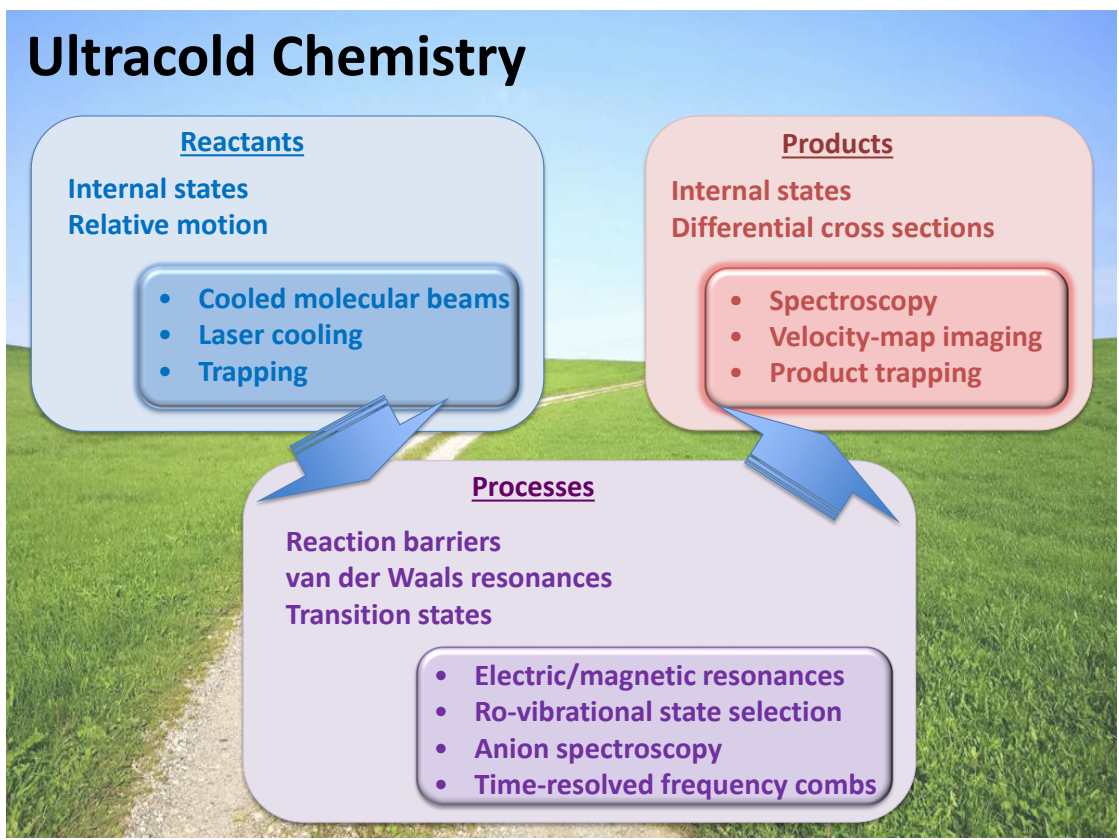

Figure 1 Landscape of cold molecule research. Cooling enables state preparation of the reactants, precise control of the reaction process, and quantitative monitoring of the transition states and products.

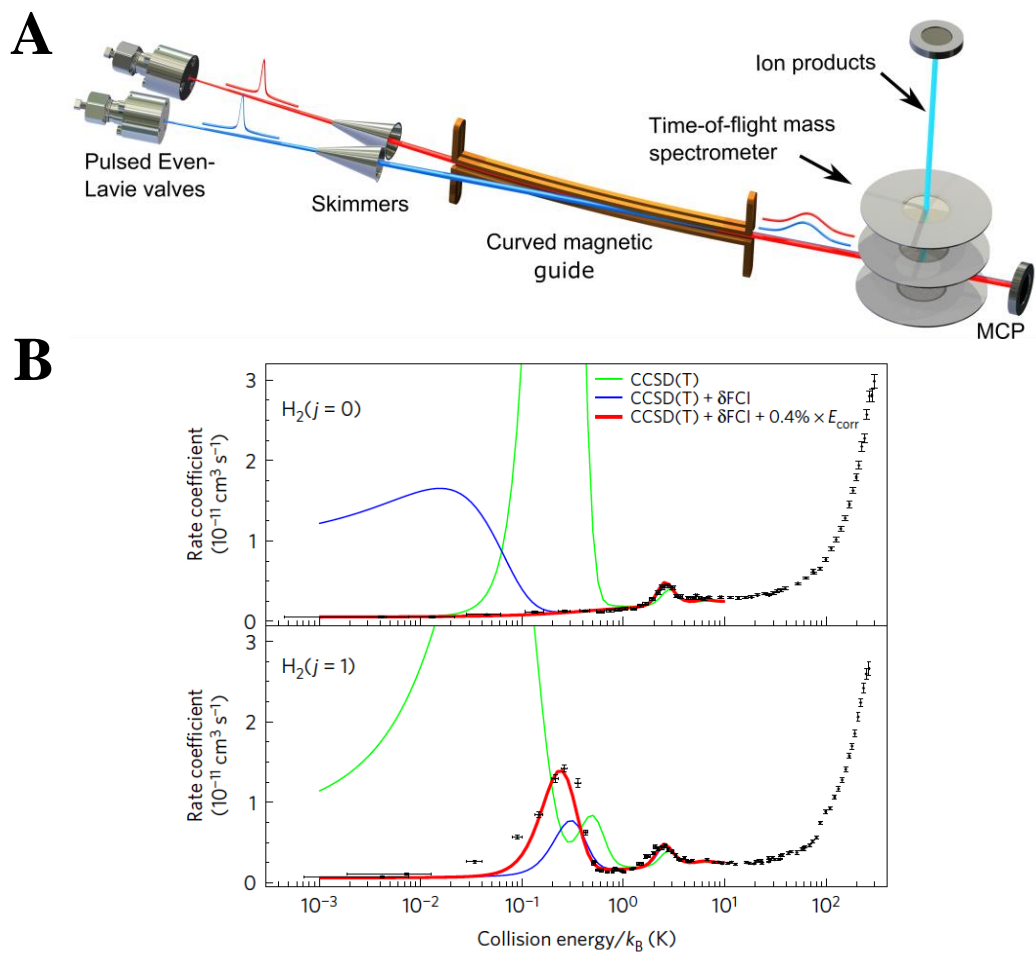

Figure 2 High resolution collision dynamics. (Top) A merged beam apparatus used for studying chemical reactions at high resolution of collision energy (below 1 K). (Middle, Bottom) Collisions of rotational-state selected $H_2$ molecules with excited (metastable) helium atoms, leading to reactive ionizations that display resonant structure as a function of collision energy. A resonance appears only if the molecule is in a rotationally excited state (Bottom), yielding a detailed probe of collision anisotropy. Figure adapted from Ref. (*50*) with permission.

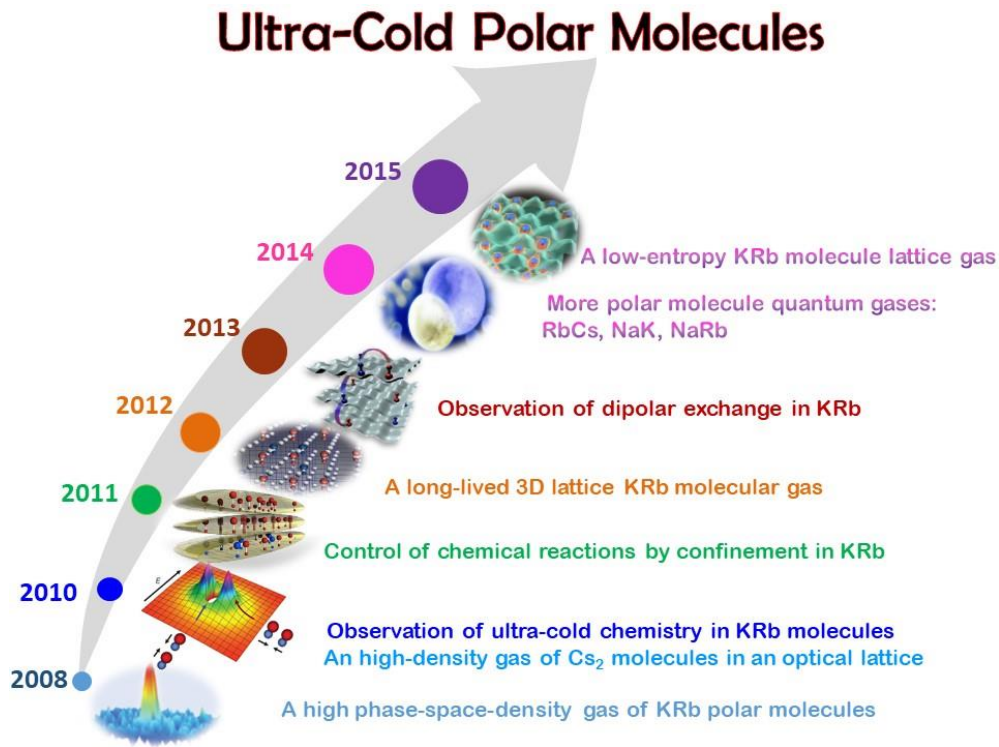

Figure 3 Experimental progress since 2008 toward production of a quantum gas of bi-alkali molecules. In chronological order: Creation of a high space-density gas of KRb polar molecules (*3*), A high-density gas of $Cs_2$ molecules in an optical lattice (*74*), Observation of ultracold chemistry in KRb molecules (*6*),Control of chemical reactions by confinement of KRb (*114*), A long-lived KRb molecular gas in a 3D lattice (*87*), Observation of dipolar exchange in KRb (*89*), More quantum gases composed of polar molecules: RbCs (*79*), NaK (*80*), NaRb (*81*), A low-entropy KRb molecular gas in a 3D lattice (*115*).

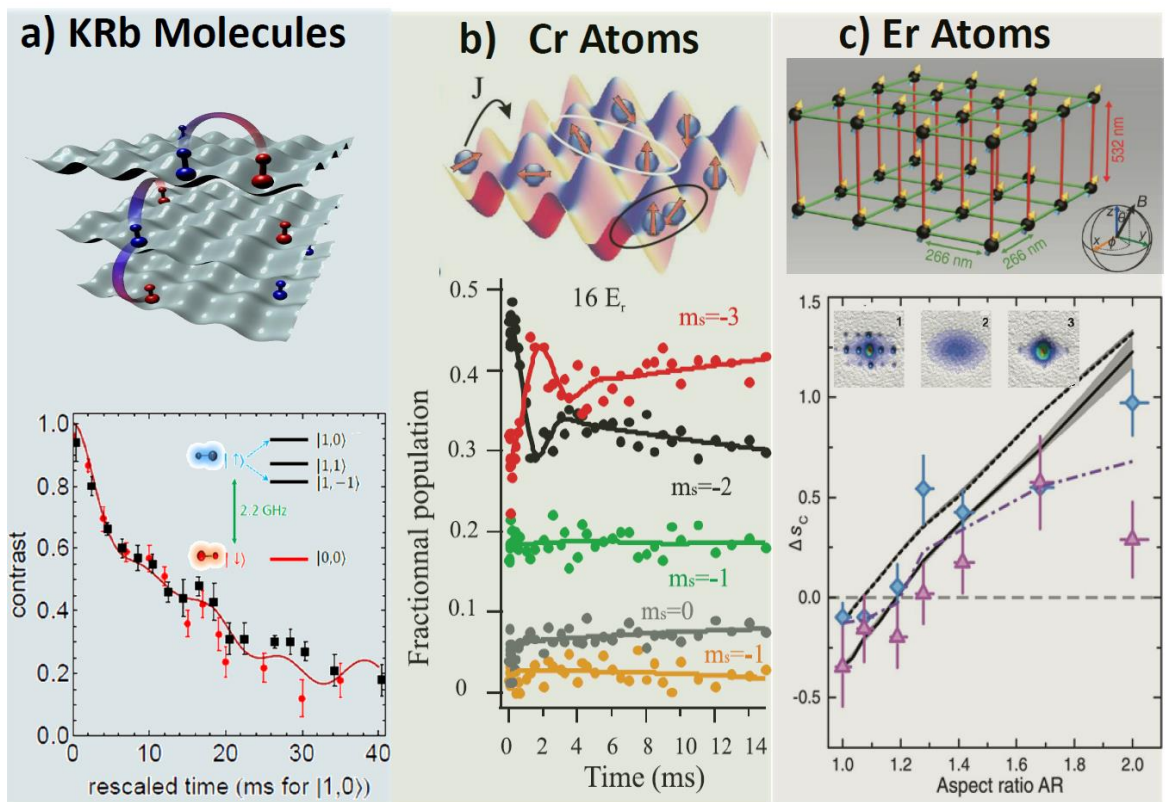

Figure 4 Experimental observation of dipolar interactions in 3D lattices using polar molecules and magnetic atoms. (Left) Measurement of Ramsey fringe contrast in a sparsely filled and deeply confining 3D lattice of KRb molecules with spins encoded in the two rotational states (*89*). (Middle) Measurement of fractional Zeeman population in a bosonic Cr gas in lattice due to dipolar interactions. Tunneling is largely suppressed in deep lattices where a Mott insulator is formed. Figure adapted from Refs. (*93, 94*) with permission. (Right) Modification of the critical lattice depth of the superfluid-to-Mott insulator transition due to the magnetic dipolar interaction between spin-polarized Er atoms. The dipolar interaction modifies the lattice depth at which the phase transition takes place in a way that depends on the atomic cloud geometry and dipole orientation. Adapted from Ref. (*95*) with permission.

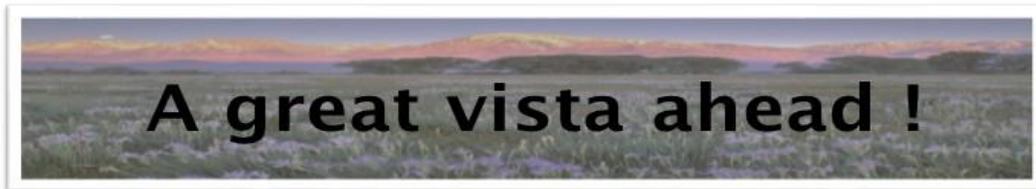

**Mobile molecules**

Topological superfluids
t-J models
Dipolar chains and clusters

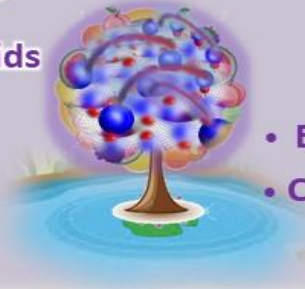

- Efficient cooling
- Control of chemistry

**Pinned multi-level molecules**

Spin-orbit coupling
Symmetry protected topological phases
Fractional Chern insulators

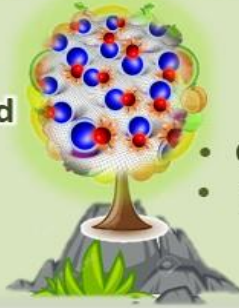

- Control of microwaves
- Local dressing

**Pinned two-level molecules**

XXZ spin model
Spin liquids
Many-body localization and transport

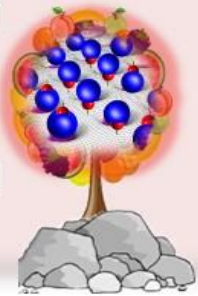

- Control of E- fields
- Higher fillings & low entropy
- Addressability

Figure 5 A range of prospective advanced quantum materials assembled with ultracold molecules.

114. We gratefully acknowledge many of our colleagues, both in and out of JILA, for their collaborations, discussions, and suggestions. We thank David Reens, John Bollinger, and Rahul Nandkishore for their comments on the paper. We acknowledge funding support for the work from ARO-MURI, AFOSR-MURI, Moore Foundation, NIST, NSF-JILA PFC-1125844, and NSF-PHY-1521080.